\documentclass[11pt]{article}
\usepackage{mymacros}
\usepackage{jheppub}

\title{Entanglement and Factorization in Axion-De Sitter Universes}
\author[a,b]{Sergio E. Aguilar-Gutierrez\orcidlink{0000-0003-0308-0061}}
\affiliation[a]{Institute for Theoretical Physics, KU Leuven,\\Celestijnenlaan 200D, B-3001 Leuven, Belgium}
\affiliation[b]{Qubits and Spacetime Unit, Okinawa Institute of Science and Technology Graduate University,\\ 1919-1 Tancha, Onna, Okinawa 904 0495, Japan}
\emailAdd{sergio.ernesto.aguilar@gmail.com}
\abstract{{We study extremal codimension-two areas and late-time bulk correlators between a pair of asymptotically de Sitter space universes connected through an Euclidean axion wormhole, in arbitrary dimensions. Assuming the validity of the de Sitter (dS)/conformal field theory (CFT) correspondence, we describe factorized Hilbert spaces for the putative boundary theories at $\mathcal{I}^+$ in each of the universes based on the asymptotically dS isometries. This allow us to we interpret the extremal areas as complex-valued holographic entanglement entropy between the microscopic duals. Later, we evaluate two-point correlation functions for heavy particles detected near $\mathcal{I}^+$. The Euclidean wormhole saddle point is responsible for finiteness of the correlators. The results are compatible with the boundary dual being non-unitary and having a large Hilbert space dimension. At last, we dimensionally reduce these geometries in terms of dilaton-gravity theory with conformally coupled matter.}}
\begin{document}

\maketitle
\newpage

\section{Introduction}
{{Entanglement entropy is a canonical measure of bipartite correlation for pure quantum states, which has a geometric counterpart in anti de Sitter (AdS)/conformal field theory (CFT) holography \cite{Maldacena:1997re} through the Ryu-Takayanagi \cite{Ryu:2006bv,Ryu:2006ef} (RT) formula (see e.g.~\cite{Nishioka:2009un,Rangamani:2016dms,Chen:2019lcd,Harlow:2014yka,Headrick:2019eth} for reviews) and its generalizations \cite{Hubeny:2007xt,Faulkner:2013ana,Engelhardt:2014gca}. This has lead to several lessons relating entanglement and spacetime connectivity. For instance, Lorentzian wormholes are associated to the entanglement entropy between the pair of CFT duals in a two-sided AdS black hole \cite{Maldacena:2013xja}. Meanwhile,}} Euclidean wormholes are central to understand the role of higher genus topology in the gravitational path integral{{, which has applications for evaluating entanglement entropy. For instance, the replica trick \cite{Lewkowycz:2013nqa} has been used to evaluate the entanglement entropy of bulk subregions in semiclassical gravity through Euclidean (replica) wormholes \cite{Penington:2019npb,Almheiri:2019psf}, which are used to prove the island formula \cite{Almheiri:2019qdq,Penington:2019kki} (in some cases including spatially disconnected universes; e.g.~\cite{Balasubramanian:2023xyd,Miyata:2021qsm,Balasubramanian:2021wgd,Balasubramanian:2020coy,Miyata:2021ncm,Iizuka:2021tut}).} {More recently,}} the role of higher genus topology in the gravitational path integral {{has been}} at the center of discussions about the one-dimensional Hilbert space of closed universes (i.e. without spatial boundaries) \cite{Harlow:2025pvj,Abdalla:2025gzn,Akers:2025ahe,Chen:2025fwp,Nomura:2025whc,Wei:2025guh,Engelhardt:2025azi,Antonini:2025ioh} (previously noted also in \cite{Almheiri:2019hni,Penington:2019kki,Balasubramanian:2023xyd}).

However, it is notoriously difficult to explicitly construct closed universes, such as de Sitter (dS) space, {{with non-trivial topological contributions, such as Euclidean wormholes, beyond lower dimensional theories and}} without introducing auxiliary spacetimes with spatial boundaries \cite{Antonini:2025ioh,Antonini:2024mci,Antonini:2023hdh}. {{An analytically solvable model where this can be realized is}} Euclidean Einstein gravity with axion fluxes and a positive cosmological constant, studied in \cite{Gutperle:2002km,Halliwell:1989pu,Myers:1988sp,klebanov1989wormholes,Aguilar-Gutierrez:2023ril} (see \cite{Chen:2016ask,Blommaert:2025bgd} for related discussions); {{and illustrated in Fig.~\ref{fig:euclidean kettle}.}} One can then construct a Friedmann– Lemaître– Robertson– Walker (FRLW) cosmology, corresponding to the solutions in \cite{Aguilar-Gutierrez:2023ril}, by cutting the (Strominger-Giddings \cite{Giddings:1987cg}-type of) wormhole in half, either at the throat, or at the location of the maximal scaling factor, {{and performing a Lorentzian continuation}} (see Fig. \ref{fig:Setup}); resulting in asymptotically dS universes connected through a quantum bounce (interpreted as a tunneling event in \cite{Aguilar-Gutierrez:2023ril}) in the presence of axion matter content. 
The latter describes an ultra-stiff fluid, which has an equation of state $\rho=p$, where $\rho$ is the energy density and $p$ is the pressure. {{As additional motivation}}, it has been recently noticed in \cite{Blommaert:2025bgd} that the axion wormhole solutions with multiple throats in \cite{Aguilar-Gutierrez:2023ril} present us with puzzles regarding their contribution to the path integral which, as remarked in \cite{Aguilar-Gutierrez:2024xfi}, are not solved by previous arguments in \cite{Halliwell:1989pu}. It is important to understand these solutions in more detail and hopefully address the issues one is confronted when taking them seriously in the path integral \cite{Halliwell:1989pu}.

{{In this work, we focus on how the presence of the wormhole seems to entangle each of the universes with its pair}}, as the Hartle-Hawking (HH) wavefunction in one of the two universes depends on the other one (through an appropriate measure of correlation between these systems). However, {{this aspect}}, hinted in \cite{Aguilar-Gutierrez:2023ril}, {{has not been made precise so far. As indicated in \cite{Aguilar-Gutierrez:2023ril}, the pair of universes have opposite arrows of time. Given that the effect of the Euclidean wormhole in this setting is very similar to the timelike wormhole in pure dS space connecting $\mathcal{I}^\pm$ when focusing on the asymptotically dS cosmologies, one might thus expect from dS/CFT \cite{Strominger:2001pn} that the correlation between the universes can be investigated in terms of entanglement entropy  between the timelike separated dual theories at $\mathcal{I}^+$ for each copy.}}
For this reason the question that we address in this work is:
\begin{quote}
    \emph{Does the wormhole entangle the pair of asymptotically dS universes it connects to?}
\end{quote}
{{We answer affirmatively the question under the assumption of asymptotically dS/CFT with a well-defined Hilbert space for the dual theories. We investigate entanglement entropy for them; based on explicit recent work in dS$_2$/CFT$_1$ \cite{Aguilar-Gutierrez:2025otq} and in analogy with results in the existing AdS/CFT literature. This shares similarities with the notions of time entanglement \cite{Narayan:2022afv,Narayan:2023ebn,Narayan:2023zen,Nanda:2025tid} and pseudoentropy in dS/CFT \cite{Doi:2022iyj}; while the details contain important differences. For instance, in our case there is no Hamiltonian evolution relating the boundary dual states; and they are treated as independent Hilbert spaces (albeit causally connected), instead of being isomorphic to each other as in \cite{Nanda:2025tid}. Either of these measures is expect to be described by the RT formula in dS/CFT \cite{Narayan:2022afv,Aguilar-Gutierrez:2025otq,Doi:2022iyj}.}} {{The computations indicate that the finite size throat of the wormhole would be encoded in the dual theory.}} 

Thus, we associate entangled dual quantum systems with a {{non-trivial Hilbert space to the holographic entanglement entropy between each of the asymptotically dS universes under assumption of dS/CFT}}. However, it has been discussed in {{different}} contexts that attaching such interpretations for pure dS space also leads to obstacles. One instance is when evaluating the late time limit of bulk correlators \cite{Dyson:2002nt}. It has been found that the correlators decay exponentially with time and reach a vanishing value asymptotically, which conflicts with the results found for generic quantum mechanical systems with a finite number of degrees of freedom that one would associate with the finite entropy of dS space {{(whose Hilbert space interpretation is debated)}}. In our work, we explore whether the asymptotically dS universes with axion fluxes in \cite{Aguilar-Gutierrez:2023ril} lead to {{similar}} type of conclusions in higher dimensions. We do the analysis in the geodesic approximation for the corresponding thermal two-point functions {{between the dual theories at $\mathcal{I}^+$}}, where one considers heavy particle states propagating along a ({{timelike}}) geodesic in the background geometry. {{The correlation function does not decay to a vanishing value at late times, although its physical interpretation is subtle since the dual theory does not need to be unitary.}}

Besides of the above results, and motivated by developments to study the information paradox \cite{Almheiri:2019psf,Penington:2019npb} in Jackiw-Teitelboim (JT) \cite{JACKIW1985343,TEITELBOIM198341} gravity; we construct a two-dimensional dilaton gravity model describing the dimensional reduction of a three-dimensional axion-dS wormholes (App. \ref{app:axion dS JT}). Here the axion particles take the role of conformally coupled matter. We denote the resulting theory axion-dS JT gravity. We hope this reduced model may also find some applications to analyze basic features of asymptotically dS spacetimes with matter, and Euclidean wormholes.

\paragraph{Structure of the manuscript}In Sec. \ref{sec:Geoemtry} we review axion-dS universes, including the Euclidean and Lorentzian solutions to the equations of motion and some analytically solvable limiting cases. In Sec. \ref{sec:entanglement} we {{study entanglement entropy between the $\mathcal{I}^+$ boundary theories of the asymptotically}} dS universes in the framework of dS/CFT holography \cite{Strominger:2001pn}. In Sec. \ref{sec:correlators} {{we evaluate bulk correlation functions between $\mathcal{I}^+$}} in each of the universes in the probe approximation. In Sec. \ref{sec:Conclusions}, we conclude with a summary of our work and some promising future directions. 

We complement the main text with appendices. For the convenience of the reader, App. \ref{app:notation} briefly summarizes the notation in the main text. {{In App.~\ref{app:correlators} we review how to evaluate correlation functions with heavy massive scalar probes through heat kernel methods.}} App.~\ref{app:axion dS JT} contains the dimensional reduction of the three-dimensional axion-dS universe, resulting in a dilaton gravity theory reproducing features of the corresponding (spatially closed) FLRW cosmology. App.~\ref{app:KSW} we find allowable metrics according to the Kontsevich-Segal-Witten criterion \cite{Kontsevich:2021dmb,Witten:2021nzp} to justify the HH-type of analytic continuation used in the main text.

\section{Brief review of axion-de Sitter wormholes}\label{sec:Geoemtry}
Axion-dS wormholes were first discussed in \cite{Myers:1988sp} and followed up in different works  \cite{Gutperle:2002km,Halliwell:1989pu,Aguilar-Gutierrez:2023ril,Aguilar-Gutierrez:2024xfi,klebanov1989wormholes,Blommaert:2025bgd} (see also \cite{klebanov1989wormholes,Chen:2025kwx}). This section reviews some general properties of the Euclidean (Sec. \ref{ssec:Euclidean}) and Lorentzian (Sec. \ref{ssec:Lorentzian}) signature solutions used in the following sections, and it sets the notation.

\subsection{Euclidean formulation}\label{ssec:Euclidean}
The $D$-dimensional theory is formulated in Euclidean signature, starting with Einstein gravity in the presence of axion matter content and a positive cosmological constant:
\begin{equation}\label{eq:onshell fundamental}
I=\int\qty[-\frac{1}{2\kappa_D^2}\star(R-2\Lambda)+\frac{1}{2}\star H_{D-1}\wedge H_{D-1}]~,
\end{equation}
where $H_{D-1}$ is the axion flux field, which is Hodge dual to the axion field $\chi$ (i.e. $H_{D-1}=\star d\chi$); $\kappa_D^2=8\pi G_N$ and we will consider a cosmological constant
\begin{equation}\label{eq:Lambda C.C.}
    \Lambda=\frac{(D-1)(D-2)}{2\ell^2}>0~.
\end{equation}
One can find spherical symmetric solutions of this theory,
\begin{align} \label{eq:metric_Euclidean_tau}
    \rmd s^2&=N(\tau)^2\rmd\tau^2+a(\tau)^2\rmd\Omega_{D-1}^2\,.
\end{align}
The axion flux field adopts the form
\begin{equation}
    H_{D-1}=Q\,
  \text{Vol}(\text{S}^{D-1})~,
  \label{eq:axion solutions}
\end{equation}
where $Q$ is a constant representing the axion flux density; and $\text{Vol}(\text{S}^{D-1})$ is the $(D-1)-$form volume element of the sphere, $\rmd\Omega_{D-1}^2$.

\begin{figure}[t!]
    \centering
    \includegraphics[width=0.35\textwidth]{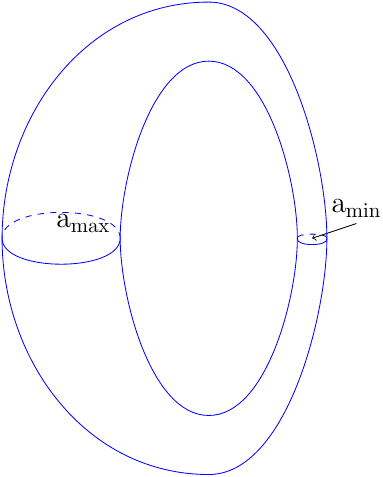}
    \caption{Euclidean geometry (in blue) of the axion-dS wormhole. The range of values of the scale factor $a\in[a_{\rm min},~a_{\rm max}]$ are determined by the roots of (\ref{eq:derivative r, tau}).}
    \label{fig:euclidean kettle}
\end{figure}

The Einstein equations then imply that the Euclidean scale factor $a(\tau)$ obeys the following constraint,
\begin{equation}
    \frac{1}{N} \frac{\rmd a}{\rmd \tau} =\pm \sqrt{1-\frac{a^2}{\ell^2} -\frac{\kappa_D^2 Q^2 \,a^{-2(D-2)}}{(D-1)(D-2)}}\label{eq:derivative r, tau}\,.
\end{equation}
We can also see that the {Euclidean} manifold for spherically symmetric axion wormholes is S$^{D-1}\times$S$^1$, which we display in Fig. \ref{fig:euclidean kettle}.\footnote{Geometrically, they similar the bra-ket wormholes in JT gravity \cite{Chen:2020tes,Milekhin:2022yzb,Fumagalli:2024msi}, yet, they describe different theories.} The argument goes as follows. One can find the locations where $\dv{a}{\tau}=0$ from the above, which indicates where the maximum and minimum values of the scale factor $a$ are located, the latter corresponds to the wormhole throat size. At these locations, we can smoothly glue the geometry by performing a periodic identification in the values of the scale factor. In principle, these solutions can be also analytically continued to describe wormholes with multiple throats \cite{Aguilar-Gutierrez:2023ril}; however, this leads to other issues \cite{Halliwell:1989pu,Blommaert:2025bgd} which we do not address in this manuscript. This means that the Euclidean time $\tau$ lives in a finite interval $\tau\in[\tau_{\rm min},\,\tau_{\rm max}]$, whose endpoints are periodically identified.
We now make a few remarks:

\paragraph{Nariai wormholes}The axion charge $Q$ cannot take arbitrarily large values\footnote{Moreover, the on-shell action of the Gibbons-Hawking instanton is smaller than for the axion-dS wormholes; and as a result, these saddles are increasingly suppressed in the Euclidean path integral as the axion charge increases.}, due to dS space being a closed universe. One can find a bound on the maximal size for the wormholes, denoted by ``\emph{Nariai wormhole}", by extremizing (\ref{eq:derivative r, tau}) with respect to $a$,
\begin{subequations}\label{eq:nariai solution}
    \begin{align}
    \kappa_D^2Q^2_{\rm max}&=\ell^{2(D-2)}(D-2)\qty(\frac{D-2}{D-1})^{D-2}\,,\label{eq:max size WH}\\
    \eval{a(\tau)}_{Q_\text{max}}&=\sqrt{\frac{D-2}{D-1}}\ell\,.\label{eq:Conf scaling factor Nariai}
\end{align}
\end{subequations}
\paragraph{Regular coordinates and the three-dimensional case}The metric (\ref{eq:metric_Euclidean_tau}) does not capture the global geometry for a generic gauge parameter $N$, given that it becomes undetermined at the $\mathcal{H}=0$ surfaces. We now look for a global coordinate system:
\begin{equation}
\rmd s^2=\rmd \tau^2+a(\tau)^2\rmd\Omega_{D-1}^2\label{eq:metric reg d} ~.
\end{equation}
The simplest case where one can find analytic solutions for $a(\tau)$ in (\ref{eq:metric reg d}), is $D=3$, where one can find
\begin{equation}\label{eq:metric in r}
    \rmd s^2=\rmd \tau^2+\frac{\ell^2}{2}\qty(1+\cos(2\frac{\tau}{\ell})\sqrt{1-\frac{2\kappa_3^2Q^2}{\ell^2}})\rmd\Omega_2^2~.
\end{equation}
Identifying the allowed range to cover the entire geometry, one identifies the periodicity of the Euclidean time in this coordinate system as $\tau\sim\tau+\pi/\ell$, {{{so that $2\abs{\tau_{\rm max}-\tau_{\rm min}}=\pi/\ell$}}.

\subsection{Lorentzian description}\label{ssec:Lorentzian}
From the Euclidean geometry, one can also learn about \emph{Lorentzian} evolution that emerges from the initial conditions specified by the HH state preparing two different universes connected through the same Euclidean saddle. A generalization of the HH proposal for this geometry was stated in \cite{Aguilar-Gutierrez:2023ril}. One slices the Euclidean geometry at either $a=a_{\rm max}$ or $a=a_{\rm min}$ and then performs the Wick rotation. The initial conditions are then determined from the Euclidean path integral preparing the state. Moreover, by a careful treatment of the scalar inhomogeneities in the background, one can find that the resulting geometries have opposite pointing arrows of time, describing bouncing universes. The resulting Lorentzian in both cases is displayed in Fig. \ref{fig:Setup}.
\begin{figure}[t!]
    \centering
\begin{subfigure}[t]{0.56\textwidth}
    \includegraphics[width=\textwidth]{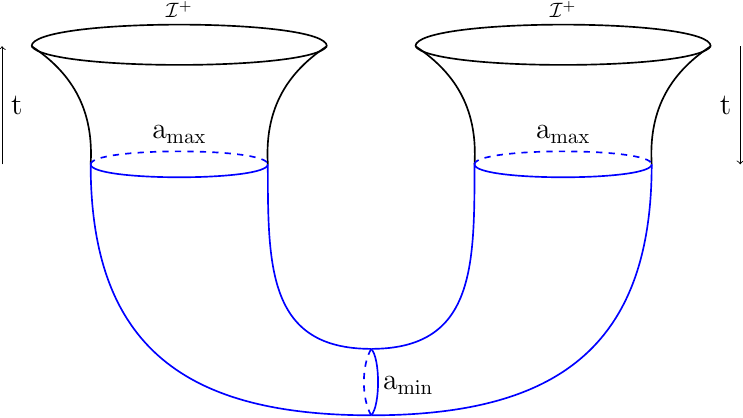}
    \caption{}
\end{subfigure}\hfill \begin{subfigure}[t]{0.42\textwidth}
    \includegraphics[width=\textwidth]{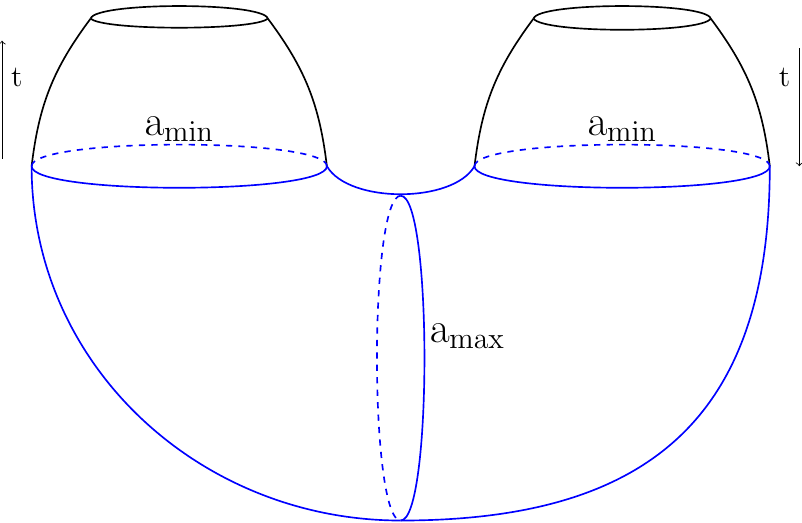}
    \caption{}
\end{subfigure}
    \caption{HH preparation of the axion-dS universes constructed by slicing the geometry twice at (a) $a=a_{\rm max}$; (b) $a=a_{\rm min}$ (shown with different relative scales), (c) both $a=a_{\rm max}$ and $a=a_{\rm min}$. The Euclidean saddle is shown with blue lines, and its Lorentzian evolution is in black, and a cyan disk bipartitioning the system. The resulting universes flow with an inverted arrow of time one from the other (black arrow), describing bouncing cosmologies.}
    \label{fig:Setup}
\end{figure}

The evolution of these universes follows from a simple Wick-rotation $\tau=\rmi t$, of \eqref{eq:derivative r, tau}, 
\begin{equation}\label{eq:coord conf gauge}
    \rmd s^2=-N(t)^2\rmd t^2+a(t)^2\rmd\Omega_{D-1}~,
\end{equation}
resulting in the equation of motion
\begin{align}
    &\qty(\frac{\dot{a}(t)}{N(t)~a(t)})^2=\frac{1}{\ell^2}-\frac{1}{a(t)^2}+\rho(t)~,\\
    &\rho:=\frac{\kappa_D^2 Q^2 a^{-2(D-1)}}{(D-1)(D-2)}~,\label{eq:axion density}
\end{align}
where $\rho$ is the energy density of the axion particles, and the dot indicates differentiation. The relation above simply represents a spatially closed FLRW cosmology; a positive cosmological constant; and axion particles, which have an equation of state $\rho=p$, the latter being the pressure.

Later, it will be useful to adopt a conformal gauge in (\ref{eq:coord conf gauge}) {{where $N(t)=a(t)$. For this gauge-fixing choice}}, $0\leq t\leq \pi/2$ with $t=0$ representing the instance in time when a quantum bounce occurs (see App. B in \cite{Aguilar-Gutierrez:2023ril}), and $-\pi/2\leq \theta_1\leq \pi/2$.
\paragraph{Disconnected geometries}
There is yet another preparation of state where we cut the axion-dS wormhole of Fig. \ref{fig:euclidean kettle}, joining $a=a_{\rm max}$ and $a=a_{\rm min}$ at the same time\footnote{We thank Mehrdad Mirbabayi for discussions on this point.} and perform the Wick rotation $\tau\rightarrow\rmi t$ to recover the corresponding quantum cosmology. The resulting geometry is illustrated in Fig. \ref{fig:4 cuts}. It corresponds to collapsing and expanding FLRW cosmologies prepared through a Euclidean wormhole. The contribution of the connected saddle point geometry to the full propagator is expected to be suppressed by a factor $\sim\rme^{I_{\rm on}^{(\rm conn)}-I_{\rm on}^{(\rm disc)}}$ with respect to disconnected saddle, where $I_{\rm on}^{(\rm disc)}=I_{\rm on}^{(\rm conn)}/2<0$ denote the on-shell actions of the disconnected and connected saddles. The on-shell actions for the axion-dS wormhole can be found in \cite{Aguilar-Gutierrez:2023ril}.
\begin{figure}[t!]
    \centering
    \includegraphics[width=\textwidth]{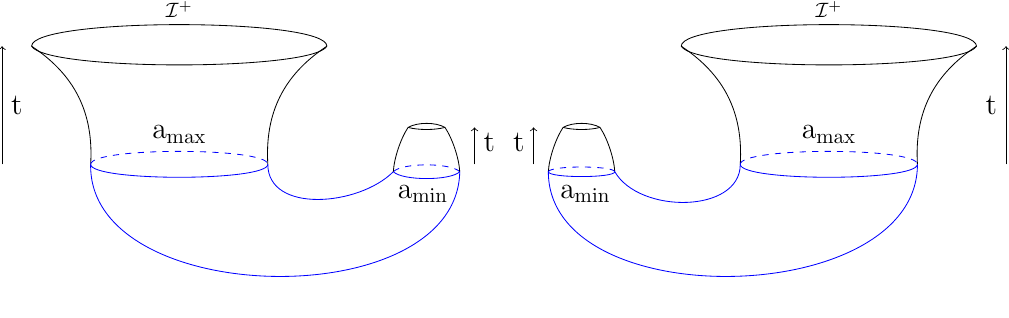}
    \caption{Four copy HH state preparation for collapsing and expanding universes, where the expanding branches are uncoupled.}
    \label{fig:4 cuts}
\end{figure}

\paragraph{Comparizon with other approaches}{{For completeness, we contrast the construction in this section with other relevant works on FLRW cosmologies resulting from analytic continuation of Euclidean wormholes. In the pioneering work \cite{VanRaamsdonk:2021qgv}, a class holographic models with features of four-dimensional big-bang/big-crunch cosmologies was constructed by analytic continuation from a pair of Euclidean CFT$_3$ coupled to a Euclidean CFT$_4$ with fewer local degrees of freedom (generating an Euclidean wormhole in the bulk from a confining three-dimensional quantum field theory in the infrared). Meanwhile, \cite{Antonini:2022ptt} considered an Euclidean asymptotically AdS$_4$ planar black hole with scalar fields dual to two Euclidean CFT$_3$ with relevant scalar operators. After analytic continuation to Lorentzian signature, one finds a cosmological spacetime a scalar potential that leads to an accelerated expansion; and where cosmological observables can be mapped from observables in the wormhole geometry. After that, \cite{Betzios:2019rds} studied asymptotically AdS wormhole saddle point solutions to a path integral and they evaluated correlation functions of local and non-local operators on distinct boundaries. They provided evidence that the boundary dual theories should factorize in the ultra-violet region. This was followed up by \cite{Betzios:2024oli}, where the authors proposed a path integral with asymptotically AdS$_4$ boundary conditions representing the wavefunction of the universe. The saddle-point solution describes Euclidean AdS half wormholes, and its Lorentzian continuation an expanding universe exhibiting (possibly long-lasting) inflation.}}

{{Thus, the FLRW cosmologies resulting from the axion-dS wormhole in this section is quite different from the previously mentioned works in that there is no AdS geometry involved. This can be less advantageous to, e.g.~discuss about entanglement in the cosmology from the boundary dual (e.g.~\cite{Antonini:2023hdh}) with a well-defined notion of a Hilbert space. Nevertheless, we argue in the next section that there are useful lessons from dS/CFT to make progress in this direction.}}

\section{Entropy from wormhole state preparation}\label{sec:entanglement}
In this section, we quantify the amount of entanglement between the dS universes that are connected through the Euclidean wormhole assuming the dS/CFT correspondence \cite{Strominger:2001pn}. {{Note that we are only concerned about the configuration in Fig.~\ref{fig:Setup} (a).}} {{We begin with Sec.~\ref{ssec:Hilbert space dual} discussing about the structure of the Hilbert space for the putative boundary theory dual based on the asymptotic dS isometries. }}Later in Sec.~\ref{ssec:holo renyi}, we evaluate the RT formula where the entangling region corresponds to $\mathcal{I}^+$ of a single one of the universes, and we comment about the holographic interpretation of the results.

\subsection{Entanglement Between Future/Past Hilbert Spaces}\label{ssec:Hilbert space dual}
{{In this section, we seek to describe entanglement between timelike separated theories in dS/CFT.
Under the assumption that there exist dual theories at $\mathcal{I}^+$ in each universe, we describe Hilbert space construction of the boundary dual (which does not need to be CFT; not even a unitary theory) and we define an associated notion of entanglement entropy.}}

{{Let us denote $\Sigma_{P/F}$ the Cauchy slices at $\mathcal{I}^+$ in each of the \emph{asymptotically dS universes}; where the $P/F$ label reflects the inverted arrow of time between the universes \cite{Aguilar-Gutierrez:2023ril} (displayed in Fig. \ref{fig:entangled Lorentzian}),  where the metric \eqref{eq:coord conf gauge} can be put in the form\footnote{{This corresponds to changing from global to static patch coordinates to describe $\mathcal{I}^+$ in the late global time limit where $\tau/\ell\gg1$ and $a(\tau)\propto \rme^{\tau/\ell}$.}}
\begin{equation}
    \eval{\rmd s^2}_{\Sigma_{P/F}}\simeq\qty(\frac{r_B}{\ell})^2\rmd \eta_{P/F}^2+r_B^2\rmd\Omega_{d-1}^2~,
\end{equation}
where $r_B/\ell\gg1$ at $\Sigma_{P/F}$; and we represent $H_{P/F}$ as the generator of the associated $\eta_{P/F}$ translations, where we may gauge fix $\eta_{P}=\eta_F\equiv\eta$.}}

{{We assume dS/CFT, meaning that there is a quantum mechanical dual operator $\hH_{P/F}$ (which is not a Hamiltonian) for boundary theories at $\Sigma_{P/F}$. Let $\ket{E_i,Q,I}_{P/F}^{(L/R)}$ represent left/right eigenstates of $\hH_{P/F}$ in the boundary dual, which is generally non-unitary:\footnote{{The dual theory is expected to be non-unitary in dS$_{d+1\geq3}$ (see explicit studies in higher spin gravity \cite{Vasiliev:1999ba,Vasiliev:1990en,Anninos:2011ui,Ng:2012xp,Anninos:2012ft,Hikida:2021ese,Hikida:2022ltr,Chen:2022ozy,Chen:2022xse}). Nevertheless, in this construction $\hH_{P/F}$ takes the role of the Hamiltonian in asymptotically AdS spacetimes.}}
\begin{equation}\label{eq:LR basis}
    \begin{aligned}
&\hH\ket{E_i,Q,I}_{P/F}^{R}=E_i\ket{E_i,Q,I}^{R}_{P/F}~,\\
    &\bra{E_i,Q,I}_{P/F}^{L}\hH=E_i\bra{E_i,Q,I}^{L}_{P/F}~,
\end{aligned}
\end{equation}
where we include the axion charge $Q$ as conserved quantum number (resulting from the shift symmetry in the field $\chi$ defined below \eqref{eq:onshell fundamental}); we also assume the basis is discrete ($i\in\qty{0,\,1,\dots,\,N}$ for some $N\in\mathbb{N}$), and $I$ denotes additional quantum numbers which depend on the complete set of commuting operators. Note that we have to differentiate left/right eigenstates since $\hH_{P/F}$ is not hermitian in general.
\\
Our working assumption is that \eqref{eq:LR basis} basis is complete in dS/CF. This is at least found in dS$_2$/CFT$_1$ \cite{Aguilar-Gutierrez:2025otq}, and it is also the analogue of the Arnowitt–Deser–Misner Hamiltonian \cite{Arnowitt:1962hi} in the asymptotically AdS case. By appropriate normalization, this basis satisfies
\begin{equation}
^L\bra{E_i,Q,I}\ket{E_j,Q,J}_{P/F}^R=\delta_{ij}\delta_{IJ}~.
\end{equation}
Similarly, we can replace $\delta_{ij}$ for $\delta(E-E')/\mu(E)$ with a corresponding $\mu(E)$ measure (which can depend on the other quantum numbers) if the spectrum is continuous in the strict $N\rightarrow\infty$ limit (and possibly involving ensemble averaging) instead.}}

{{We define the Hilbert space of the boundary theories on each hypersurface $\Sigma_{P/F}$ for a fixed axion charge as
\begin{equation}
    \mathcal{H}^{(Q)}_{F/P} \equiv{\rm span}\qty{\ket{E_i,Q,I}_{F/P}^R,~\bra{E_i,Q,I}_{F/P}^L}~.
\end{equation}
We have depicted the spacetime and its holographic interpretation in Fig. \ref{fig:entangled Lorentzian} using the coordinates (\ref{eq:coord conf gauge}) in the conformal gauge $N(t)=a(t)$.}}
\begin{figure}[t!]
    \centering
    \includegraphics[width=\textwidth]{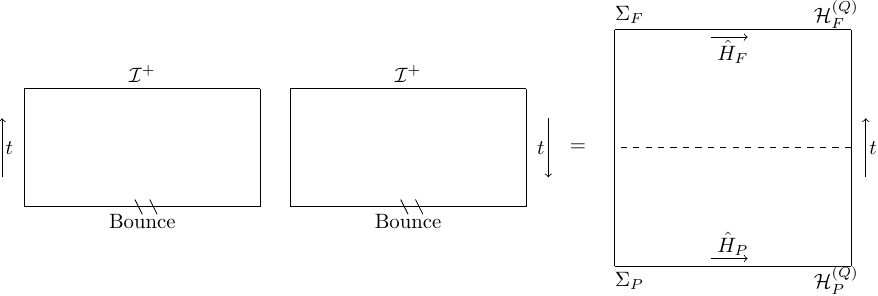}
    \caption{{Spatially closed FLRW universes with an ultra-stiff fluid evolving between bounce to a inflating cosmologies coupled at $t=0$ though the Euclidean wormhole. Given that the arrows of time evolve in opposite directions, the system can be represented in a single Penrose diagram, where $\Sigma_{P/F}$ are the Cauchy surfaces where the Hilbert space $\mathcal{H}_{P/F}^{(Q)}$ of the boundary theories are defined, and $\hH_{P/F}$ is the generator of translations along $\eta_{P/F}$.}}
    \label{fig:entangled Lorentzian}
\end{figure}
{{We can then construct entangled states from $\mathcal{H}_{P/F}$ such as
\begin{equation}\label{eq:EE}
\begin{aligned}
    \ket{\Psi}^R:=\sum_{ijIJ}\psi^R_{ijIJ}\ket{E_i,Q,I}^R_P\otimes\ket{E_j,Q,J}^R_F~,\\
\ket{\Psi}^L:=\sum_{ijIJ}\psi^L_{ijIJ}\ket{E_i,Q,I}^R_P\otimes\ket{E_j,Q,J}^R_F~,
\end{aligned}
\end{equation}
where $\psi^{L/R}_{ijIJ}$ are, in principle, independent coefficients. If $\hH_{P/F}=\hH_{P/F}^\dagger$, then $\psi^{R}_{ijIJ}=(\psi^{L}_{ijIJ})^*$; however, we will not make this assumption.\\
\\
In particular we can have a past/future analog of the thermofield double (TFD)-like state when
\begin{equation}\label{eq:TFD}
    {\rm TFD}~:\quad\psi^L_{ijIJ}=\psi^R_{ijIJ}=\frac{\rme^{-\beta E_i/2}}{\sqrt{Z(\beta)}}\delta_{ij}\delta_{IJ}~,\quad Z(\beta):=\sum_j\rme^{-\beta E_j}~.
\end{equation}
Note that 
\begin{enumerate}[label=\roman*]
    \item We do not consider a superposition of states with different values of $Q$, where we would include a chemical potential term in the TFD-like state, since in the bulk we only describe axion wormhole solutions with fixed $Q$.
    \item $Z(\beta)$ above might not be interpreted as a partition function since $\hH_{P/F}$ is generator of spatial translations; instead of a Hamiltonian.
\end{enumerate}
\paragraph{Time entangled states}We now move to study entanglement between the states at $\Sigma_{F/P}$. We first construct normalized density matrices corresponding to a generic pure (left-)state $\ket{\Psi}$ in \eqref{eq:EE}, which is generated by the two-sided HH state preparation, as
\begin{equation}
    \rho_{\Psi}:=\frac{\ket{\Psi}^R\bra{\Psi}^L}{\bra{\Psi}^L\ket{\Psi}^R}~.
\end{equation}
We trace out either past or future subsystems to construct reduced density matrices respect to the remaining subsystem, e.g.
\begin{equation}\label{eq:reduced F}
    \begin{aligned}
        &\tr_P(\cdot):=\sum_{I}\qty(\bra{E_i,Q,I}^L_P\otimes\mathbb{1})\cdot\qty(\ket{E_i,Q,I}^R_P\otimes\mathbb{1})\\
        \rho_{F}:=&\tr_P(\rho_{\Psi})=\sum_{ijkIJK}\psi_{ijIJ}^R\psi^L_{kjKJ}\ket{E_i,Q,I}_F\bra{E_k,Q,K}_F~,
    \end{aligned}
\end{equation}
where we take $\sum_{ijIJ}\abs{\psi_{ijIJ}}^2=1$ as the normalization.}}

{{One can see from this construction that there can be a non-trivial von Neumann entropy
\begin{equation}
    S=-\Tr_F(\rho_F\log\rho_F):=\sum_{l}\bra{E_l,Q}\rho_F\log\rho_F\ket{E_l,Q}~,
\end{equation}
for the reduced density matrix \eqref{eq:reduced F} for the past/future entangled states \eqref{eq:EE} when the corresponding state preparation results in non-factorised states (e.g.~$\ket{\Psi}=\ket{E}_P\otimes\ket{E}_F$). For instance, in the thermofield double-like state \eqref{eq:TFD}, one finds that
\begin{equation}
    S=\log Z(\beta)-\beta\partial_\beta \log Z(\beta)~.
\end{equation}
In contrast to the Bekenstein-Hawking entropy in pure dS, the von Neumann entropy above might not have a statistical interpretation as thermodynamic entropy respect to an observer since there is no properly a Hamiltonian associated to the theories at $\mathcal{I}^+$, and $Z(\beta)$ \eqref{eq:reduced F} does not need to represent a partition function.}}

\subsection{Holographic Entanglement from Future/Past Slices}\label{ssec:holo renyi}
{{Based on the working assumption of dS/CFT there should be a geometrical description for the entanglement entropy of the timelike separated boundary theories. A natural expectation, based on AdS/CFT and an explicit lower dimensional example in dS$_2$/CFT$_1$ \cite{Aguilar-Gutierrez:2025otq}, which we study in the next subsection, is that the holographic dual to the quantity defined in this section is given by an extremal area according to the RT formula. 
\begin{equation}\label{eq:Renyi entropy}
    S=\frac{A(\gamma)}{4G_N}~,
\end{equation}
where $A(\gamma)$ is the area of the minimal codimension-two surface of a extremal surface $\gamma$ homologous to the entangling region at $\mathcal{I}^+$ in  Fig. \ref{fig:Setup} (a) for {{a pair}} of universes (left, or right), and we take to be S$^{D-1}$ for the geometry in (\ref{eq:metric reg d}). To show the validity of this assumption, one would need a explicit boundary theory dual for the gravitational theory; which is outside the scope of this work. Nevertheless, we hope it contributes towards a geometric dual of entanglement entropy between the theories at $\mathcal{I}^+$.}}

 {{Note that in contrast, Fig.}}~\ref{fig:Setup} (b) corresponds to a collapsing universe, which does not have dS asymptotic conditions. We will only analyze the universes that are asymptotically dS with the entangling surfaces at $\mathcal{I}^+$, which we take to be S$^{D-1}$ surfaces{{, and connected through the Euclidean wormhole at $t=0$. We now proceed with the evaluation of \eqref{eq:Renyi entropy} for the different components of Fig.~\ref{fig:Setup} (a).}}

\paragraph{Lorentzian section}Let us first express
\begin{equation}\label{eq:metric regular}
    \rmd\Omega_{D-1}^2=\rmd\theta_1^2+\cos^2\theta_1\rmd\theta_2+\dots+\cos^2\theta_1\dots\cos^2\theta_{D-2}\rmd \theta_{D-1}^2~,
\end{equation}
{\textcolor{blue}{in}} \eqref{eq:coord conf gauge}{\textcolor{blue}{ t}}o specify the RT surface in terms of the location of the entangling surface{\textcolor{blue}{ at}} $\mathcal{I}^+$ in (\ref{eq:metric regular}), where the {{codimension-two}} area functional {{for the Lorentzian region for $D\geq4$}} in the two-sided HH preparation of state then reads
\begin{equation}\label{eq:Area functional}
    A^{\rm (L)}(\gamma)=2\int\rmd\Omega_{D-3}\int\rmd\chi\sqrt{-\qty(\dv{t}{\chi})^2+a(t)^2\qty(\dv{\theta_1}{\chi})^2}a(t)^{D-3}\qty(\cos\theta_1)^{D-3}~,
\end{equation}
where $\chi$ is an arbitrary parametrization of the codimension-two surface, {\textcolor{blue}{we have taken $\theta_{D-1}=0$ as part of the ansatz,}} {{and the factor 2 accounts for the contribution from both axion-dS universes. This evaluation of RT surfaces from $\mathcal{I}^+$ is in agreement with the known pure dS$_D$ result (when $Q=0$) in \cite{Narayan:2022afv}; although the boundary interpretation in the previous section is different from theirs even in the $Q=0$ case (since we do not define the Hilbert spaces from Hamiltonian eigenstates).}}  

{{Next,}} the extremal condition $\delta A^{(L)}(\gamma)=0$ in \eqref{eq:Area functional} requires that $\theta_1=$constant, {{which is fixed by the anchoring surface. Let $\theta_1=\theta_0$ and $t=t_c\gg\ell$, which acts a regulator, as the boundary conditions at $\mathcal{I}^+$. Then, the contribution of the codimension-two Euclidean extremal surface is just
\begin{equation}\label{eq:cosmic brane I}
    A^{\rm (L)}(\gamma)=2\rmi\cos^{D-3}\theta_0~\Omega_{D-3}\int_{0}^{t_c}\rmd t ~a(t)^{D-3}~,
\end{equation}
where $\Omega_{D}=\frac{2\pi^{D/2}}{\Gamma\qty(\frac{D}{2})}$. An analytically solvable case for \eqref{eq:cosmic brane I} in any $D$ is the Nariai limit in (\ref{eq:Conf scaling factor Nariai}), where we get
\begin{equation}\label{eq:AL Nariai}
    \eval{A^{\rm (L)}(\gamma)}_{Q\sim Q_{\rm max}}\simeq2\rmi t_c\cos^{D-3}\theta_0~\Omega_{D-3}\qty(\frac{D-2}{D-1})^{\frac{D-3}{2}}\ell^{D-3}~.
\end{equation}}}
Notice however, that the case $Q=Q_{\rm max}$ is the Einstein static universe, which is not asymptotically dS space, so (\ref{eq:Entropy dS CFT}) does not apply in the limiting case, although the result might still provide an appropriate approximate answer for $Q\sim Q_{\rm max}$ {{at sufficiently late times}}.

{{Meanwhile, in $D=3$, the ansatz \eqref{eq:Area functional} for the codimension-two extremal surface is modified to
\begin{equation}\label{eq:AL D3}
    \eval{A^{\rm (L)}(\gamma)}_{D=3}=2\rmi\int_0^{t_c}\rmd t=2\rmi t_c~.
\end{equation}
\paragraph{Euclidean section}We now move to the Euclidean evaluation, where the functional \eqref{eq:Area functional} (resulting from continuation $\tau\rightarrow\rmi t$) reduces to
\begin{equation}\label{eq:Entropy dS CFT}
   A^{(\rm E)}(\gamma)=\begin{cases}
       2\Omega_{D-3}\int\rmd\chi\sqrt{\qty(\dv{\tau}{\chi})^2+a(\tau)^2\qty(\dv{\theta_1}{\chi})^2}a(\tau)^{D-3}\qty(\cos\theta_1)^{D-3}~,&D\geq4\\
       2\abs{\tau_{\rm min}-\tau_{\rm max}}~,&D=3~.
   \end{cases}
\end{equation}
Note that while in the Lorentzian section there were Dirichlet boundary conditions (at $\mathcal{I}^+$ for each universe) in the functional after gauge-fixing the coordinate system along geometry connecting the entangling surfaces \eqref{eq:Area functional}; in the Euclidean section, we are using continuity along the interface between Lorentzian and Euclidean geometries at $t=\tau=0$ where $\theta_1=$constant for $\delta A^{\rm (E)}=0$ in \eqref{eq:Entropy dS CFT} (i.e.~the boundary condition in the integral follow from continuity of the RT surface).\footnote{{{This similarly happens with single entangling region at $\mathcal{I}^+$ in the dS case with a no-boundary surface (joining the Lorentzian dS hyperboloid with the Euclidean hemisphere) \cite{Doi:2022iyj,Narayan:2022afv}.}}}}} 

{{(\ref{eq:Entropy dS CFT}) in the $Q\simeq Q_{\rm max}$ and $D=3$ cases therefore results in the \eqref{eq:AL Nariai} and \eqref{eq:AL D3} respectively under the substitution $\rmi t_c\rightarrow\abs{\tau_{\rm max}-\tau_{\rm min}}$. 
\paragraph{Timelike Entanglement} Taking the previous results together in \eqref{eq:Renyi entropy}, we have that
\begin{subequations}\label{eq:enttropies all}
\begin{align}
        \eval{S(\gamma)}_{Q\sim Q_{\rm max}}&\simeq\frac{\Omega_{D-3}}{2G_N}(\rmi t_c+\Delta\tau)(\cos\theta_0)^{D-3}\qty(\frac{D-2}{D-1})^{\frac{D-3}{2}}\ell^{D-3}~,\\
\eval{S(\gamma)}_{D=3}&=\frac{\rmi t_c+\Delta\tau}{2G_N}~.
\end{align}
\end{subequations}
These cases serve to illustrate that the boundary dual of the pair of universes are entangled in time due to the presence of the Euclidean wormhole coupling them. In particular, they are complex-valued and require a regularization parameter $t_c$, as found in the pure dS$_{d+1\geq3}$ cases (e.g. \cite{Narayan:2022afv,Doi:2022iyj}). This motivates to study differences and similarities with eternal AdS black holes \cite{Maldacena:2001kr} in the next subsection.}}

\section{Late-time correlators}\label{sec:correlators}
{{\eqref{eq:enttropies all} suggests that the associated dual theory might not be unitary \cite{Doi:2022iyj,Narayan:2022afv}, which is allowed from the perspective of the dual Hilbert space construction in Sec.~\ref{ssec:Hilbert space dual}. In contrast when the theory is unitary and the entropy finite (such as the Gibbons Hawking entropy \cite{Gibbons:1977mu}), there can be an interpretation in terms of a finite dimensional Hilbert space (which in principle requires a sum over topologies in the gravitational path integral). In fact, it is expected that correlation functions must fluctuate at late proper times for unitary chaotic systems with a finite number of degrees of freedom \cite{Mirbabayi:2023vgl}, leading to a non-vanishing norm for the late-time correlator functions. To make contrast with the AdS/CFT case, we investigate the evolution of late time correlation functions between the pair of universes in this section, although one should not expect that the results are similar for the aforementioned reasons.}} 

In Sec. \ref{ssec:geodesics} we study the geodesic joining the $\mathcal{I}^+$ boundary theories in Fig. \ref{fig:Setup} (a); while in Sec. \ref{ssec:connected} we evaluate the correlation functions between the corresponding surfaces.

\subsection{Geodesic approximation}\label{ssec:geodesics}
We now add a massive free scalar field, $\phi$, in the axion-dS theory (Fig. \ref{fig:Setup}), where the scalar theory is given:
\begin{equation}\label{eq:scalar action}
    I_{\rm scalar}=-\frac{1}{2}\int\rmd^{D} x\sqrt{-g}~\qty(\qty(\partial\phi)^2+m^2\phi^2)~.
\end{equation}
We can then compute its propagator in different ways. We will employ the geodesic approximation, valid when $\phi$ is a heavy massive field \cite{Balasubramanian:1999zv,Louko:2000tp} {{to solve the Klein-Gordon equation of the scalar field \eqref{eq:scalar action} using the heat kernel method in App. \ref{app:correlators} (see \eqref{eq:large mass limit}),\footnote{{{Given that we work in a closed universe, some simplifications for propagators in Euclidean AdS wormholes (e.g.~\cite{Betzios:2019rds}) do not apply in this case, but the evaluation follows from the same principles.}}}
\begin{equation}\label{eq: G geodesics}
    G(x,~y)\simeq-\rmi\sqrt{\Delta(x,y)}\frac{\qty(m~L(x,y))^{\frac{D-3}{2}}}{2^{D-\frac{1}{2}}\pi^{(D-1)/2}}\rme^{-m~L(x,y)}~,
\end{equation}
where $L(x,~y)$ is the geodesic length between the points $x$, $y$; $\Delta(x,y)$ is the Van Vleck-Morette determinant shown in \eqref{eq:important eq}. Note that (\ref{eq: G geodesics}) becomes more accurate when the points $x$ and $y$ are close to each other, or when $m$ is large compared to the inverse geodesic length connecting $x$ and $y$ \cite{Aalsma:2022eru,Chapman:2022mqd}.}}

\subsection{Correlators in Connected Expanding Universes}\label{ssec:connected}
Next, study the late-time correlator according to the dual theories close to $\mathcal{I}^+$ in each of the universes shown in Fig \ref{fig:correlators} with a propagator passing through the wormhole geometry.
\begin{figure}[t!]
    \centering
    \includegraphics[width=0.75\textwidth]{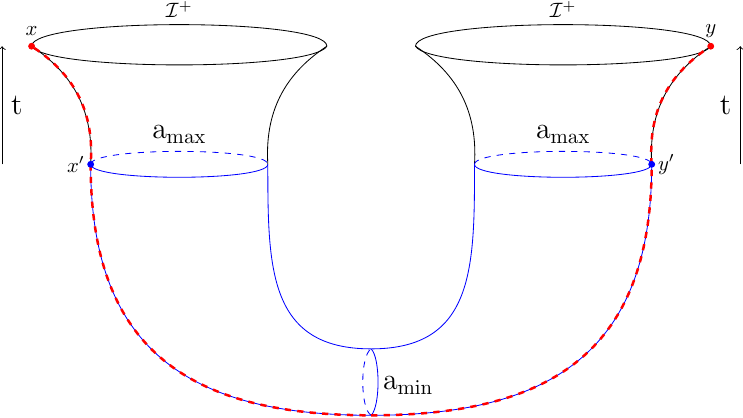}
    \caption{Correlator for very heavy fields following a geodesic (red dashed line) between points $x$ and $y$ (red dots) very near $\mathcal{I}^+$ on each axion-dS universe, and they are connected by a geodesic path to $x'$, $y'$ (blue dots) to the Euclidean wormhole.}
    \label{fig:correlators}
\end{figure}
Let us denote the Lorentzian contributions to the geodesic length with $L^{\rm (g)}(z_i,~z_i')$, and $z_i=\qty{x,~y}$, $z'_i=\qty{x',~y'}$, {{and the Euclidean part as $L^{\rm (g)}(x',~y')$.
\begin{equation}\label{eq:total geodesic}
    L^{(g)}(x,y)=L^{(g)}(x,x')+L^{(g)}(y,y')+L^{(g)}(x',y')~.
\end{equation}}}
To specify the region of evaluation, we consider that the points $x$, $y$ in the ``twin" universes, are located in the coordinate $t$, $\qty{\theta_i}$ (\ref{eq:metric reg d}) in patches where the arrow of time is inverted with respect to each other (see Sec. \ref{ssec:Lorentzian}). The geodesic equation requires the same for the points $x'$, $y'$ in the $a=a_{\rm max}$ slice of the Euclidean wormholes.

We start the evaluation with the geodesics connecting $x$ to $x'$, and $y$ to $y'$ in Fig \ref{fig:correlators}. Considering the Wick rotation $\tau\rightarrow\rmi t$ in (\ref{eq:metric reg d}), so we can specify the geodesic length in these patches as
\begin{equation}\label{eq:geo length}
    L^{\rm (g)}(z_i,~z_i')=\int_{\lambda_{z_i}}^{\lambda_{z_i'}}\rmd\lambda~\sqrt{-\qty(\dv{t}{\lambda})^2+a(t(\lambda))^2\qty(\dv{\Omega_{D-1}}{\lambda})^2}~,
\end{equation}
where $\lambda$ is a general parametrization for the geodesic. 

Since we implement spatially symmetric points for $x$, $y$, as well $x'$ and $y$, we have that $\lambda_{x}=\lambda_{y}$ and $\lambda_{x'}=\lambda_{y'}$, {{i.e.
\begin{equation}
    L^{\rm (g)}(x,~x')=L^{\rm (g)}(y,~y')~,
\end{equation}
which are complex-valued each. This means,}} we are considering a configuration with exact symmetry in terms of the operator insertions at $x$ and $y$ between the universes of Fig. \ref{fig:correlators}, mostly for technical simplicity. However, this is perhaps the most natural configuration to consider given the similarity to the late-time correlators between antipodal observers in pure dS space, which shows quasi-periodic decaying behavior \cite{Aalsma:2022eru}. However, it would be interesting to consider how the late-time correlation functions would be modified for more general configurations.

To make simplifications and carry out the calculation, we set the endpoints of the geodesics at $\theta_i(\tau=\tau_{\rm max}) = 0$ $\forall~1\leq i\leq D-1
$. Using the geodesic equations of motion in \eqref{eq:geo length}
\begin{equation}
    \pdv{L^{(g)}}{\theta_i}=\partial_\tau\pdv{L^{(g)}}{\dot{\theta}_i}~,
\end{equation}
one can verify that 
\begin{equation}\label{eq:geodesic path}
\theta_i^{\rm (g)}=0,    \quad 1\leq i\leq D-1~,
\end{equation}
is the only solution to the geodesic equations respecting the boundary conditions. {{Using the boundary conditions $t=t_c$ at $\mathcal{I}^+$ for each universe, we have that
\begin{equation}
     L^{\rm (g)}(x,~x')=\rmi t_c~.
\end{equation}}}
Meanwhile, the Euclidean contribution is given by $L^{\rm (g)}(x',~y')$ in Fig.~\ref{fig:correlators}. To account for the Euclidean section that the geodesic curve traverses, we set $\tau$ as the parametrization of the geodesic path, such that (\ref{eq:metric reg d}) has a corresponding length functional with \eqref{eq:geodesic path}
    \begin{equation}\label{eq:geodesic Euclidean}
    \begin{aligned}
        L^{\rm (g)}(x',~y')&=\int_{\tau_{\rm min}}^{\tau_{\rm max}}\rmd\tau~\sqrt{1+a(\tau)^2\qty(\dv{\Omega_{D-1}}{\tau})^2}=2\Delta\tau~,
    \end{aligned}
\end{equation}
where $\Delta\tau\equiv\abs{\tau_{\rm max}-\tau_{\rm min}}$ is just the periodicity of the wormhole. 
{{\paragraph{Feynman Green's Function}
Adding up the previous results into the complete geodesic \eqref{eq: G geodesics}, the squared norm of the Feynman Green's function \eqref{eq: G geodesics} becomes
\begin{equation}\label{eq:correlation final}
   \abs{G^{(\rm conn)} (x,y)}^2={\Delta(x,y)}\frac{m^{D-3} \qty(\Delta\tau^2+ t_c^2)^{\frac{D-3}{2}}}{4\pi^{D-1}}\rme^{-2m\Delta\tau}~,
\end{equation}
where, again, $\Delta(x,y)$ is shown in \eqref{eq:important eq}. This indicates that $\abs{G^{(\rm conn)} (x,y)}^2$ does not vanish at $t_c/\ell\gg1$, in contrast to the eternal AdS black hole case \cite{Maldacena:2001kr}, where the correlation function (without additional non-trivial topologies) vanishes at late times. The difference is that geodesics connecting the boundaries are timelike instead of spacelike (Sec. 4.3). As emphasized at the start of the section, one should not expect that the correlation function \eqref{eq:correlation final} would saturate to a finite value; which is what one would expect for a unitary theory consistent with finite degrees for freedom (of the order $\rme^{S}$) leading to a late time erratic behavior in the $\abs{G}^2$. Instead, \eqref{eq:correlation final} is consistent with the evaluation of the entropy \eqref{eq:enttropies all} in indicating that the dual theory maybe non-unitary (due to the contribution of timelike and spacelike geodesics to the correlator) and it has possibly infinite degrees of freedom (indicated by the cutoff parameter $t_c$). Similar to \cite{Maldacena:2001kr} the dimensionality of the degrees of freedom in the dual theory might be reduced once we allow for higher genus topologies in the path integral of the theory \eqref{eq:onshell fundamental}.}}

\section{Conclusions}\label{sec:Conclusions}
{{Let us first summarize the work. We studied the entangled between the boundary theories at $\mathcal{I}^+$ for a pair or FLRW universes with positive cosmological constant and axion matter having late time dS asymptotics. We argued from dS/CFT and the late time dS isometries about the structure of the Hilbert space of the boundary theory dual, and how to define entanglement entropy between them. Using dS/CFT principles, we evaluated the entanglement entropy holographically through the RT formula with $\mathcal{I}^+$ of each universe as the anchoring regions for the extremal surfaces. Here, the wormhole throat size and the timelike codimension-two extremal areas determines the corresponding holographic entanglement entropy.}}

{{To explore similarities with the eternal AdS black hole, we also described a pair of decoupled and entangled boundary theories duals, we calculated the magnitude of the late-time two-point correlator functions with respect to $\mathcal{I}^+$ of the universes, considering connected saddles Fig. \ref{fig:correlators}. We showed that, in the geodesic approximation of the magnitude of the two-point correlation functions does not decay exponentially in Lorentzian time, and it is compatible with the complex holographic entanglement entropy. The Euclidean wormhole connecting the universes allows a non-vanishing late-time correlation function connecting the universes, that reflects expected properties of the (non-unitary) holographic dual theory.}}

\subsection{Outlook}
We now discuss some future directions related to this work.

{{First, a crucial direction to justify the assumptions made in this work is to find a explicit boundary dual to the dS wormhole; to show that there is an interpretation for the quantities that we defined and the correspondence between the entanglement entropy for temporal separated theories with the extremal area in the RT formula.}} For this purpose, it might be useful to adopt some of the explicit approaches to dS/CFT space and to find a way of including axion matter, such as higher spin gravity \cite{Vasiliev:1999ba,Vasiliev:1990en} in dS$_4$ space \cite{Anninos:2011ui,Ng:2012xp,Anninos:2012ft} or a SU($2$) Wess-Zumino-Witten model dual to dS$_3$ space \cite{Hikida:2021ese,Hikida:2022ltr}.

{{Second, we have focused on entanglement entropy between boundary theories at $\mathcal{I}^+$. It would be interesting to recast our study in terms of bulk operator algebras in dS/CFT. One could rigorously study the type of von Neumann algebra corresponding to this system, which is claimed to be type I in \cite{Blommaert:2025bgd} (although not based on the dS/CFT perspective in this work). To do that, we should elaborate on the Hilbert space construction of the boundary theories, and construct an algebra of observables acting on them, such as from our study of matter correlation functions in Sec.~\ref{sec:correlators}. It would be useful to find an explicit boundary dual theory to expand on the Hilbert space construction and the notion of partial traces and reduced density matrices (Sec.~\ref{ssec:Hilbert space dual}) from concrete algebraic arguments.}}

{{Third, one could explore the effects of the axion wormholes in the entanglement entropy and correlation functions for the dual CFTs in a the AdS analog of our work (e.g. \cite{Gutperle:2002km}) where one can rely on ordinary AdS/CFT}}. However, it has been noticed that there are possible inconsistencies with the holographic dictionary if the Euclidean wormholes end on multiple boundaries \cite{Hertog:2017owm,Loges:2023ypl}, and the Lorentzian interpretation remains more unclear than the setting we have been considering.

At last, the fact that the universes are entangled suggests that should be some redundant information encoded between them. This is similar to false vacuum decay eternal inflation in quantum cosmology, where there is manifest redundancy in the description of the theory producing different identical universes \cite{Hartle:2016tpo}. It would be natural to investigate if there is a notion of quantum error correction between the pair of universes from dS/CFT. One might want to start generalizing the Hamilton-Kabat-Lifschytz-Lowe reconstruction \cite{Hamilton:2005ju,Hamilton:2006fh,Hamilton:2006az,Hamilton:2007wj} for this setting.

\section*{Acknowledgements}
I thank Andreas Blommaert, Thomas Hertog, Joel Karlsson, Mehrdad Mirbabayi, Edward K. Morvan, Jan Pieter van der Schaar, and Thomas Van Riet for very useful conversations. SEAG thanks the organizers of the ``Cosmology, Quantum Gravity, and Holography: the Interplay of Fundamental Concepts" CERN workshop for their hospitality during the early stages of this work; as well as Amsterdam University, the Delta Institute for Theoretical Physics, the Research Foundation - Flanders (FWO), and the International Centre for Theoretical Physics for travel during different phases of the project.
The work of SEAG was partially supported by the FWO Research Project G0H9318N and the inter-university project iBOF/21/084. SEAG is supported by the Okinawa Institute of Science and Technology Graduate University. This project/publication was also made possible through the support of the ID\#62312 grant from the John Templeton Foundation, as part of the ‘The Quantum Information Structure of Spacetime’ Project (QISS), as well as Grant ID\# 62423 from the John Templeton Foundation. The opinions expressed in this project/publication are those of the author(s) and do not necessarily reflect the views of the John Templeton Foundation.

\appendix

\section{Notation}\label{app:notation}
\paragraph{Acronyms}
\begin{itemize}[noitemsep]
    \item (A)dS: Anti-de Sitter
	\item CFT: Conformal field theory
    \item FLRW: Friedmann–Lemaître–Robertson–Walker 
    \item HH: Hartle-Hawking
    \item JT: Jackiw-Teitelboim
    \item KS: Kontsevich-Segal
\end{itemize}
\paragraph{Definitions}
\begin{itemize}
    \item $\Lambda=(D-1)(D-2)/2\ell^2$ \eqref{eq:Lambda C.C.}: cosmological constant.
    \item $a(\tau)$, $N(\tau)$ \eqref{eq:metric_Euclidean_tau}: scale factor and the lapse respectively as function of the Euclidean time.
\item $H_{D-1}$, $Q$ \eqref{eq:axion solutions}: Axion flux field, and axion charge density respectively.
\item $Q_{\rm max}$, $a_{\rm max}$, $a_{\rm min}$ \eqref{eq:nariai solution}: Maximal charge, maximal and minimal scale factors for the Nariai wormhole limit.
\item $\rho$ \eqref{eq:axion density}: axion matter density.
\item $\mathcal{H}_{P/F}$: Boundary theory states at $\mathcal{I}^+$ in asymptotically dS FLRW cosmologies.
\item $\hH_{P/F}$: Dual operator to the generator of spatial translations at $\mathcal{I}^+$.
\item $\ket{E,Q}_{P/F}$: Eigenstates in $\mathcal{H}_{P/F}$ for a wormhole with fixed axion charge density $Q$.
\item $S$ \eqref{eq:Entropy dS CFT}: Entanglement entropy.
\item $G(x,~y)$ \eqref{eq: G geodesics}: Two-point correlation function in the geodesic approximation.
\item $L^{\rm (g)}(x,y)$ \eqref{eq:geo length}: Geodesic lengths between spacetime points $x$ and $y$.
\end{itemize}
{{\section{Correlation functions}\label{app:correlators}
In this appendix, we review the procedure to evaluate the Feynman Green's function in generic curved spacetimes by solving the KG equation from the heat kernel method \cite{Vassilevich:2003xt}. The Green’s function of the scalar $\phi$ in \eqref{eq:scalar action}, $G(x,y):=\expval{\phi(x)\phi(y)}$ satisfies
\begin{equation}\label{eq:EOM G}
    (\Box-m^2)G=\frac{\delta^{D}(x-y)}{\sqrt{-g(x)}}~.
\end{equation}
\paragraph{Hear kernel method}The above expression can be solved using the heat kernel method, where we express
\cite{Bekenstein:1981xe}
\begin{equation}\label{eq:propagator}
    G(x,y)=:\rmi\int\rmd s~K(x,y;s)~\rme^{-\epsilon s}~.
\end{equation}
Here the factor $\rme^{-\epsilon s}$ is for convergence of the integral (which makes $G$ a Feynman Green's function), so that we will be taking $\epsilon\rightarrow0$ in the evaluation; and $K(x,y;s)$ obeys the Schrodinger equation
\begin{equation}
    (\square-m^2)K(x,y;s)=-\rmi \partial_sK(x,y;s)~,
\end{equation}
where $\square$ acts on either the $x$ or $y$ coordinates, and there is a boundary condition
\begin{equation}
    \lim_{s\rightarrow0}K(x,y;s)=\frac{\delta^{D}(x-y)}{\sqrt{-g(x)}}~,
\end{equation}
that follows from \eqref{eq:EOM G}. The heat kernel solution in general curved spacetime is given by \cite{Vassilevich:2003xt}
\begin{equation}\label{eq:heat general}
    K(x,y;s)=-\rmi\frac{\rme^{\rmi m^2 s+\frac{\rmi \sigma(x,y)}{2s}}}{(4\pi\rmi s)^{D/2}}\sqrt{\Delta(x,y)}(1+\mathcal{O}(s))~,
\end{equation}
where $s$ is the parametrization of the geodesic; with $s\ll1$ corresponding to $x$ and $y$ being close to each other, corresponding to a large mass of the scalar $m$ with respect to the other length scales \cite{Vassilevich:2003xt}; and 
\begin{subequations}\label{eq:VV}
\begin{align}
    \sigma(x,y):&=\frac{1}{2}\qty(L(x,y))^2~,\\
    \Delta(x,y):&=\frac{1}{\sqrt{g(x)g(y)}}\det\qty(\pdv{\sigma(x,y)}{x^\nu}{y^\mu})~,
\end{align}
\end{subequations}
where the latter is the Van Vleck-Morette determinant, which represents the leading correction to saddle point approximation of the scalar $\phi(x)$ in the Green's function $G(x,y)$. Note that the above solution corresponds to a single geodesic curve approximation; while in general there can be multiple geodesics connecting points $x$ and $y$ in a general spacetime. The Feynman propagator \eqref{eq:propagator} can then be evaluated from the heat kernel \eqref{eq:heat general} and the determinant \eqref{eq:VV} as \cite{Vassilevich:2003xt}
\begin{equation}\label{eq:Feynmann evalauted}
    G(x,y)\simeq-\rmi\sqrt{\Delta(x,y)}\frac{\qty(m\sqrt{2\sigma(x,y)})^{\frac{D-2}{2}}}{2^{d}\pi^{D/2}}K_{\frac{D-2}{2}}\qty(m\sqrt{2\sigma(x,y)})~,
\end{equation}
where $K_n(\cdot)$ is the modified Bessel function second kind. Note that the dependence on the specific curved spacetime background is encoded only in $L(x,y)$ and $\Delta(x,y)$. In particular, in the large mass limit, the asymptotic behavior of \eqref{eq:Feynmann evalauted} becomes
\begin{equation}\label{eq:large mass limit}
    G(x,y)\sim-\rmi\sqrt{\Delta(x,y)}\frac{\qty(m\sqrt{2\sigma(x,y)})^{\frac{D-3}{2}}}{2^{D-\frac{1}{2}}\pi^{(D-1)/2}}\rme^{-m\sqrt{2\sigma(x,y)}}~,
\end{equation}
which is expected from saddle point approximation for the Green's function \cite{Chapman:2022mqd}.
\paragraph{Evaluating the determinant}We can proceed evaluating the determinant $\Delta$. It is known that for general spacetimes, it satisfies \cite{Decanini:2005gt,Kent:2013ouo}:
\begin{equation}\label{eq:EOM VDM}
    \frac{\partial_\sigma\sqrt{\Delta}}{\sqrt{\Delta}}=-\frac{\square_x(\sqrt{2\sigma})}{2\sqrt{2\sigma}}+\frac{d-1}{4\sigma}~,
\end{equation}
which is subject to a boundary condition
\begin{equation}\label{eq:bdry cond Delta}
    \eval{\Delta(x,y)}_{\sigma=0}=1~.
\end{equation}
We evaluate the Laplacian in the geodesics length using similar methods as \cite{Allen:1985wd} (see discussion below (1.11)). An important difference is that we work in a wormhole or FLRW background instead of a maximally symmetric space; so that $\square_x L(x,y)$ (where $L=\sqrt{2\sigma}$) is not generically maximally symmetry. However, we are particularly interesting in geodesics where $L=\tau$ in the coordinate system \eqref{eq:metric_Euclidean_tau}, which simplifies the evaluation,
\begin{equation}
    \square_x L(x,y)=(D-1)\frac{a'(\tau)}{a(\tau)}~.
\end{equation}
This means that for geodesics where $L=\tau$ \eqref{eq:EOM VDM} becomes
\begin{equation}\label{eq:important eq}
    \sqrt{\Delta}=\exp(\int_0^\sigma\rmd\sigma'\qty[\frac{D-1}{4\sigma'}-\frac{1}{2\sqrt{2\sigma'}}\frac{a'(\sqrt{2\sigma})}{a(\sqrt{2\sigma})}])~.
\end{equation}
At this point we have not yet specified $a(\tau)$.\footnote{{{In the dS$_{d+1}$ case (i.e. $Q=0$), where $a(\tau)=\ell\sin(\tau/\ell)$, one finds a known in \cite{Aalsma:2022eru}.}}} We can explore \eqref{eq:important eq} further using the $D=3$ metric \eqref{eq:metric in r}, where 
\begin{equation}
    \eval{\frac{a'(\tau)}{a(\tau)}}_{D=3}=\frac{2}{\ell}\frac{\sqrt{1-\mu^2}\sin(\frac{2\tau}{\ell})}{1-\sqrt{1-\mu^2}\cos(\frac{2\tau}{\ell})}~.
\end{equation}
where $\mu:=\sqrt{2}\kappa_3Q/\ell\in[0,1]$; so that \eqref{eq:important eq} becomes
\begin{equation}
    \eval{\Delta(x,y)}_{D=3}=\mathcal{N}\frac{\sigma(x,y)}{\qty(1-\sqrt{1-\mu^2}\cos(2\sqrt{2\sigma(x,y)}))^2}~,
\end{equation}
where $\mathcal{N}$ is a regularization constant for the initial condition \eqref{eq:bdry cond Delta} to be satisfied.\footnote{{{Given that $\mathcal{N}$ is an overal regularization constant, the determinant remains well defined even though it requires that $\mathcal{N}\rightarrow\infty$ for $\Delta(x,x)=1$. Without the overall regularization constant, the heat kernel method would fail to describe correlation functions as soon as $\mu\neq0$. However, the asymptotic behavior described by \eqref{eq:large mass limit} is in agreement with the geodesic approximation in App. \ref{sapp:saddle point geodesic} which does not require regularization for the one-loop corrections to the geodesic approximation.}}}}}

{{Similarly, for the Nariai limiting case \eqref{eq:Conf scaling factor Nariai} where $a(\tau)$ is constant \eqref{eq:nariai solution}, we similarly recover $\Delta(x,y)_{Q_\text{max}}=\mathcal{N}\sigma^{(D-1)/2}$.
with $\mathcal{N}$ a regularization constant.}}
{{
\subsection{Saddle point approximation}\label{sapp:saddle point geodesic}
We test an alternative evaluation of the correlation function in the previous subsection using a saddle point approximation for the path integral of the Green's function \eqref{eq:EOM G} of the scalar field \eqref{eq:scalar action} in general spacetimes \cite{Chapman:2022mqd} in 
\begin{equation}
    G(x,y)=\int\rmd[P]~\rme^{-m~L[P]}~,
\end{equation}
where $P$ represents different paths connecting points $x$ and $y$.}}
The corresponding propagator at tree-level in the geodesic approximation (\ref{eq: G geodesics}) is then 
\begin{equation}
    G^{\rm (conn)}(x,~y)\sim \rme^{-2m\abs{\tau_{\rm max}-\tau_{\rm min}}}~.
\end{equation}
One can also add one-loop corrections for the evaluation of (\ref{eq: G geodesics}) as outlined in \cite{Chapman:2022mqd}. We will study small deviations from the geodesic path (\ref{eq:geodesic path}) considering $\theta_i=\theta_i^{\rm (g)}+\delta \theta_i(\tau)$ with $\delta \theta_i(\tau)\ll1$. 
In this case (\ref{eq:geodesic Euclidean}) has a corresponding second-order variation
\begin{equation}\label{eq:1-loop corrected}
    \delta L=\int\rmd\tau~\frac{a(\tau)^2}{2}\sum_{i=1}^{D-1} \qty(\dv{\delta\theta_i}{\tau})^2~.
\end{equation}
Then, the one-loop corrections to (\ref{eq: G geodesics}) can be evaluated through path integral methods
\begin{equation}\label{eq:path int 1-loop}
\begin{aligned}
    G^{\rm (conn)}(x,~y)=\sum_{g\in\text{geodesics}}\rme^{-{m}~L^{\rm (g)}}\int\prod_{i=1}^{D-1}[d\delta\theta_i]\rme^{-m~\delta L}~.
\end{aligned}
\end{equation}
Combining (\ref{eq:1-loop corrected}) and (\ref{eq:path int 1-loop}), one can use the path integral evaluation in App. C of \cite{Chapman:2022mqd} for the geodesic lengths to derive that
\begin{equation}
    \begin{aligned}\label{eq:G with 1-l correction}
    G^{\rm (conn)}(x,~y)&= \rme^{-m\Delta\tau}\sqrt{\frac{m}{2\pi f(\tau_{\rm min},~\tau_{\rm max})}}~,\\
    f(\tau_{\rm min},~\tau_{\rm max})&=\int_{\tau_{\rm min}}^{\tau_{\rm max}}\frac{\rmd\tau}{a(\tau)^2}~.
\end{aligned}
\end{equation}
In principle, the analytic evaluation of (\ref{eq:G with 1-l correction}) for a generic wormhole can be done numerically for a scale factor $a(\tau)$ in regular coordinates; and it can be done analytically for the $D=3$ case in (\ref{eq:metric in r}). For higher dimensions, we can illustrate the result by working in the regime where $Q\sim Q_{\rm max}$, such that $a(\tau)$ is approximately the constant (\ref{eq:Conf scaling factor Nariai}). In that case (\ref{eq:G with 1-l correction}) becomes
\begin{equation}\label{eq:pure Euclidean}
   \eval{G^{\rm (conn)}(x,~y)}_{Q\sim Q_{\rm max}}= \rme^{-m\Delta\tau}\sqrt{\frac{2}{\pi}\frac{(D-2)m\ell^2}{(D-1) \Delta\tau}}~.
\end{equation}
{{This provides an alternative evaluation, albeit less precise than the heat kernel answer in \eqref{eq:large mass limit}.}}

\section{Axion-dS JT gravity}\label{app:axion dS JT}
In this appendix, we deduce a two-dimensional dilaton gravity theory, which we call axion-dS JT gravity, by dimensionally reducing the axion-dS model (\ref{eq:onshell fundamental}) in three-dimensions. In the following, $D=3$ quantities are denoted with hats, and two-dimensional ones are unhatted and have Latin indices. 

We begin with the Euclidean theory (\ref{eq:onshell fundamental}) in $D=3$
\begin{equation}
    I_{\rm bulk}=-\int\rmd^3x\sqrt{\hat{g}}\qty[\frac{1}{2\kappa_3^2}(\hat{\mathcal{R}}-2\Lambda)+\frac{1}{2}(\hat{\nabla}\chi)^2]~,
\end{equation}
and we consider an ansatz of the form
\begin{equation}
    \rmd s^2=g_{ij}\rmd x^i\rmd x^j+\Phi^{2}\rmd \varphi^2~.
\end{equation}
Using the metric in (\ref{eq:metric in r}), the dilaton above can be identified as:
\begin{equation}\label{eq:Dilaton from dS EWH}
    \Phi(\tau,\,\theta)=a(\tau)\sin\theta~.
\end{equation}
The effective action becomes
\begin{equation}\label{eq:1-loop EFT action}
    I_{\rm bulk}=-2\pi\int\rmd^2x\sqrt{g}\Phi\qty[\frac{1}{2\kappa_3^2}(\mathcal{R}^{(2)}-2\Lambda)+\rho]~,
\end{equation}
where $\rho:=\frac{1}{2}(\hat{\nabla}\chi)^2=\frac{Q^2}{2a^4}$ is the energy density of the (massless) axion field.

The equations of motion for the lower dimensional system (\ref{eq:1-loop EFT action}) are
\begin{equation}
\begin{aligned}
    &R=2\Lambda+16\pi G_3\rho~,\\
    -&(\nabla_i\nabla_j-g_{ij}\nabla^2)\Phi+\Lambda g_{ij}\Phi=8\pi G_3\Phi T_{ij}~,
\end{aligned}
\end{equation}
which agree with (\ref{eq:derivative r, tau}). 

Although we have focused on the Euclidean theory; the continuation to the Lorentzian signature simply describes a spatially closed FLRW cosmology, positive cosmological constant, and an equation of state $p=\rho$, with a corresponding stress tensor
\begin{equation}
    T_{ij}=pg_{ij}+(\rho+p)u_i u_j~.
\end{equation}
The resulting 2D theory is a special case of a dilaton gravity theory where axion particles are conformally coupled matter. A family of two-dimensional FLRW gravity models with similar characteristics have been previously studied in JT gravity \cite{Cadoni:2002pu}.

\section{Kontsevich-Segal criterion \& Schwinger-Keldysh-like contour}\label{app:KSW}
In this appendix, we provide new results on the allowable complex metrics interpolating between the Euclidean wormholes (\ref{eq:metric_Euclidean_tau}) and the expanding quantum cosmology (\ref{eq:coord conf gauge}) in the HH construction. See \cite{Blommaert:2025bgd,Halliwell:1989pu} for related discussions about the contour in the complex plane for the axion-dS wormholes.

The Kontsevich-Segal (KS) criterion \cite{Kontsevich:2021dmb} (see also \cite{Witten:2021nzp}) states that complex metrics on a manifold is allowable if it leads to a convergent path integral for all $p$-form matter fields of arbitrary quantum field theory. The condition can be stated in terms of eigenvalues, $\lambda_i$, of the metric, $g_{\mu\nu}$, as \cite{Kontsevich:2021dmb}:
\begin{equation}\label{eq:KSW crit}
    \sum_i\abs{\text{arg}(\lambda_i)}<\pi~.
\end{equation}
We now search an explicit family of KS-obeying solutions with a complexified ``time-like'' coordinate $\tau$ in (\ref{eq:metric_Euclidean_tau}), which is bounded by the marginally violating curves (i.e. where the sum of arguments in (\ref{eq:KSW crit}) gives $\pi$). Let $\mathcal{C}$ and $\mathcal{C}'$ denote curves in the $\tau\in\mathbb{C}$ plane where (\ref{eq:KSW crit}) is marginally violated, and $\chi$ is a general parametrization of the curves in the complex $\tau$ plane. This curve must then satisfy:
\begin{equation}\label{eq:marginal curve}
    \abs{\text{arg}(N(\tau)^2~\tau'(\chi)^2)}+(D-1)\abs{\text{arg}(a(\tau)^2)}=\pi~,
\end{equation}
which can then be expressed as
\begin{equation}
    \dv{\tau}{\chi}=\frac{\rmi}{N(\tau)a(\tau)^{D-1}}~.
\end{equation}
Using (\ref{eq:derivative r, tau}) we can determine the curve:
\begin{equation}\label{eq:minimally violating curve}
\rmi~\chi=\int\rmd\tau~N(\tau)a(\tau)^{D-1}=\int\rmd a\qty(1-\frac{a^2}{\ell^2}-\frac{\kappa_D^2a^{-2(D-2)}}{(D-1)(D-2)})^{-1/2}{a^{D-1}}~.
\end{equation}
For example, we may consider $D=3$ in \eqref{eq:minimally violating curve} to simplify the expressions, where we find:
\begin{equation}\label{eq:chi tau}
    \chi=\frac{\sqrt{2}\ell}{4}\qty(-\sqrt{\tilde{\Delta}(\tau)}+\ell^2\arctan\frac{2a(\tau)^2-\ell^2}{\sqrt{\tilde{\Delta}(\tau)}})~,\quad \tilde{\Delta}(\tau):={2a(\tau)^2\ell^2-\kappa_D^2\ell^2-2a(\tau)^4}>0~.
\end{equation}
{{By inverting \eqref{eq:chi tau}, we recover the curve $\mathcal{C}'$ in Fig. \ref{fig:KSW curves}; while $\mathcal{C}$ corresponds to the two-sided HH preparation in (\ref{eq:metric_Euclidean_tau}, \ref{eq:coord conf gauge}.)}} The family of curves {{interpolating between them}} are schematically shown in Fig. \ref{fig:KSW curves}. {{Note that these curves are reminiscent of the Schwinger-Keldysh contour \cite{Schwinger:1951ex,Keldysh:1964ud}, where the forward and backwards-evolving branches represent the pair of universes with inverted arrow of time. The case analyzed in the main text corresponds to the curve $\mathcal{C}$ in Fig.~\ref{fig:KSW curves}.}}
\begin{figure}
    \centering
    \includegraphics[width=0.7\textwidth]{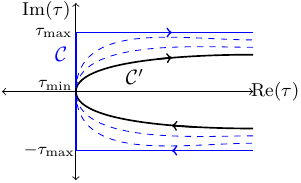}
    \caption{Examples of contours satisfying the KS criterion (dashed blue curve) and those that marginally violate it, including the doubled HH continuation between the Euclidean wormholes (\ref{eq:metric_Euclidean_tau}) {{and the Lorentzian universes (\ref{eq:coord conf gauge}) (blue solid curve, $\mathcal{C}$), and (\ref{eq:chi tau}) (solid black curve, $\mathcal{C}'$). In the main text we take $\tau_{\rm max}$ and $\tau_{\rm max}$ in the opposite way with respect to this figure (i.e. $\tau_{\rm max}=0$ to do the analytic continuation at $\tau=t=0$ and $\tau_{\rm min>0}$) which was selected for clarity in the representation.}}}
    \label{fig:KSW curves}
\end{figure}

\bibliographystyle{JHEP}
\bibliography{references.bib}
\end{document}